\documentclass{aa}

\usepackage[varg]{txfonts}
\usepackage{graphicx}
\usepackage{xcolor}
\usepackage[inline]{enumitem}
\usepackage{tabularx}

\usepackage{natbib}
\bibpunct{(}{)}{;}{a}{}{,}

\usepackage{hyperref}
\hypersetup{colorlinks=true,citecolor=blue,linkcolor=blue}

\begin{document}

\title{
    From planetesimals to planets with $N$-body simulations in the giant-planet formation region
    }

\author{
    Sebastian~Lorek \and
    Michiel~Lambrechts
    }

\institute{
    Center for Star and Planet Formation,
    Globe Institute, 
    University of Copenhagen,
    {\O}ster Voldgade 5–7, 
    DK-1350 Copenhagen, 
    Denmark \\
    \email{sebastian.lorek@sund.ku.dk}
    }

\date{Received ; accepted }

\abstract{
The cores of wide-orbit giant planets can form via pebble accretion if large planetesimals form in the outer regions of protoplanetary discs at sufficiently early times. Streaming instability simulations support mass distributions consistent with Solar System minor body constraints, but when and where planetesimal formation took place remains uncertain. Here, we report on our $N$-body simulations of core formation through pebble and planetesimal accretion starting from streaming-instability inspired planetesimal mass distributions. We explore two initial radial planetesimal distributions, a ring-like and a spatially more uniform distribution, between $10$ and $50\,\mathrm{AU}$. To address the numerical challenge of simulating realistic planetesimal numbers, corresponding to one to ten Earth masses of planetesimals, we made use of GPU acceleration for the $N$-body interactions (with \texttt{GENGA}) and a newly developed pebble accretion module. We find that the top of the planetesimal mass distribution provides the seeds for core formation through pebble accretion, leading to the formation of multiple giant planets. This is consistent with previous studies not including $N$-body interactions. Planetesimal surface densities, crudely corresponding to an initial $10\,\%$ formation efficiency, imply low mean collision rates (around unity) in the gas disc phase. Our simulations show that giant planet formation depends only weakly on the initial locations where planetesimals form, because of rapid dynamical scattering, and on their total mass budget, due to filtering of the pebble flux between embryos. After disc dissipation, giant planet systems stir the remnant primordial planetesimals, making a scattered disc an inherent outcome of giant planet formation. Giant impacts between planetary cores generally appear to be rare in the first $100\,\mathrm{Myr}$.
}

\keywords{Methods: numerical -- Planets and satellites: formation}

\titlerunning{From planetesimals to planets in the giant-planet formation region}
\maketitle
\nolinenumbers

\section{Introduction}
\label{sec:introduction}

In the core accretion scenario, giant planets form in the protoplanetary disc by first consolidating a solid core with a mass of around ten Earth masses, after which a $\mathrm{H}${/}$\mathrm{He}$ envelope is accreted \citep{Mizuno1980,Pollack1996}. Such young gaseous protoplanetary discs contain a large mass reservoir of millimetre- to centimetre-sized icy dust particles called pebbles. Observations hint that discs around solar-mass stars frequently start out with pebble mass budgets in the range of tens to hundreds of Earth masses \citep{Appelgren2023,Appelgren2025}. 

These pebbles are the likely building blocks of the first planetesimals, bodies larger than ten to one hundred kilometres in size. In the disc midplane, a dense pebble layer can become unstable to the streaming instability \citep{Youdin2005}. Subsequently, filaments form that then collapse by self-gravity into planetesimals \citep{Johansen2007}, with a mass distribution consistent with that of the minor body populations of the Solar System \citep{Johansen2015,Simon2016,Schaefer2017}. 

Currently, it is unclear how efficient pebbles are converted into planetesimals. Historically, models for giant-planet core growth assumed high surface densities of planetesimals. However, such formation models only considering planetesimals still face challenges to forming giant planets within gas disc lifetimes \citep{Pollack1996,Thommes2003,Coleman2014}. For example, accretion cross-sections are low for primordial planetesimals larger than ten kilometres in size \citep{Rafikov2004}, embryos deplete planetesimal feeding zones \citep{Levison2010}, and the migration of planetary cores may lead to planetesimal shepherding instead of accretion \citep{Tanaka1999,JohansenBitsch2019}.

In contrast, if planetesimal formation is less efficient, the remaining pebble mass fraction can be swept up by planetesimals, thus accelerating core growth. Small particles are efficiently accreted: Pebbles passing the planet spiral to the core due to gas drag and allow the planet to sweep up approximately 10\,\% of the inwards drifting pebble flux \citep{Ormel2010,Lambrechts2012}. In the outer regions of the disc, where stellar irradiation dominates heating, this leads to inside-out growth of embryos \citep{Lambrechts2012,Ida2016}. This allows cores to grow to completion under nominal pebble flux and pebble size ranges for solar metallicities and above \citep{Lambrechts2014b,Bitsch2015,Savvidou2023,Gurrutxaga2024}. 

The $N$-body simulations that include pebble accretion show that a form of oligarchic growth, where only a handful of cores emerge, can be supported when large embryos stir neighbouring planetesimals out of the pebble midplane \citep{Levison2015}. However, these simulations ignore type-I migration. Subsequent $N$-body studies of giant-planet formation, including planetary migration, revealed a wide variety of planetary system architectures at disc dissipation, which later settle to stable multi-planet systems after the dissipation of the gas disc. \citep{Matsumura2017,Bitsch2019,Wimarsson2020}. Such $N$-body studies are valuable to initiate now in order to begin a comparison with the growing population of giant planets in wide orbits \citep{Bitsch2020b,Matsumura2021}. Such cold giants are now routinely detected through long-term radial velocity surveys and direct imaging. The occurrence rate of giant planets (${\gtrsim}0.1$\,Jupiter mass), located between $0.3$ and $30\,\mathrm{AU}$, is around $30\,\%$ \citep{Fulton2021}. At wider orbits, direct imaging places a limit on the occurrence rate of around $5\,\%$ of super-Jupiters \citep{Nielsen2019, Vigan2021}. The multiplicity of giant exoplanets is not well known. Kepler data support warm giants typically being in multiple systems \citep{Mulders2018}, and this is supported by radial velocity surveys as well \citep{Mayor2011}. At wider orbits, moderate eccentricities ($e{=}0.2{-}0.6$) are common, hinting that these planets are also frequently located in multiples, although this may partly be a detection bias \citep{Rosenthal2021,Rosenthal2024}.

In this work, we aim to investigate the growth from planetesimals to planets throughout the giant-planet formation region while taking pebble accretion into account. This is numerically challenging, and therefore previous works have frequently assumed the existence of planetary embryos to reduce the number of simulated particles. However, in order to understand the critical transition from the planetesimal formation and growth phase to the phase where embryos are sufficiently massive, it is important to have their growth dominated by pebble accretion \citep{Liu2019}, which sensitively depends on the assumed planetesimal mass distribution and surface density \citep{Lorek2022}. The objective of this work is not to perform exoplanet population synthesis \citep{Matsumura2021} nor to explore the formation pathway of the Solar System \citep{Raorane2024}. Such explorations would require larger simulation suites to explore stochastic variations and different protoplanetary disc conditions, such as the spread in their initial solid mass budgets, disc lifetimes, or heating mechanism.

It is not clear where the initial population of planetesimals formed, so we assumed a standard uniform distribution similar to \citet{Lau2024} but also one where planetesimals form in spatially separated rings. Because observed discs frequently show over- and under-dense pebble rings \citep{Andrews2020}, it has been proposed that planetesimal formation may similarly be spatially non-uniform \citep{Coleman2016}. We, however, do not assume that these rings are associated with pressure bumps and do not explicitly model these in the gas disc \citep{Morbidelli2020,Brasser2020,Lau2022,Izidoro2022,Jiang2023}.

Tracing the full planetesimal population will also allow us to understand directly what happens to the primordial population of planetesimals after planet formation has been completed. The formation of giant planets creates a scattered disc of minor bodies, which is a process that took place in the Solar System and is believed to be linked to the frequently observed debris discs around young stars \citep{Wyatt2020}.

The paper is organised as follows. In Sect.~\ref{sec:methods} we provide an overview of the $N$-body method, the physical background model, the pebble accretion prescription, and the initial conditions for our simulations. In Sect.~\ref{sec:results} we discuss the simulation results. In Sect.~\ref{sec:discussion} we put our work into context. And finally, in Sect.~\ref{sec:conclusions}, we summarise the main results.

\section{Methods}
\label{sec:methods}

In this section we outline the methods used in our $N$-body simulations of planet formation. In Sect.~\ref{sec:discmodel} we explain our choice for the protoplanetary disc. In Sect.~\ref{sec:pebbleaccretion} we detail the pebble accretion prescription, the choice of Stokes number, and the reduction of the pebble flux through accreting planets. In Sect.~\ref{sec:gasaccretion} we provide the equations for the gas accretion prescription we use to model the stage past the pebble isolation mass. In Sect.~\ref{sec:planetesimalsizes} we introduce the streaming-instability initial-mass function that determines the initial size distribution of the planetesimals in our simulations. In Sect.~\ref{sec:gpunbodysimulations} we present the $N$-body method we chose for the simulations. Finally, in Sect.~\ref{sec:initialconditions} we outline the initial conditions for our study. A summary of the parameters and their values used in our simulations is provided in Table~\ref{tab:initialconditions}.
\begin{table}
\caption{Simulation parameters and values.}
\label{tab:initialconditions}
\centering
\begin{tabularx}{\columnwidth}{lll>{\raggedright\arraybackslash}X}
\hline\hline
parameter & value & unit & notes \\
\hline
$M_\star$ & $1$ & $M_\odot$ & stellar mass \\
$t_0$ & $0.3$ & $\mathrm{Myr}$ & initial time \\
$t_\mathrm{neb}$ & $4.1$ & $\mathrm{Myr}$ & final time nebular phase \\
$t_\mathrm{gf}$ & $100$ & $\mathrm{Myr}$ & final time gas-free phase \\
$\Delta t$ & $12$ & $\mathrm{days}$ & time step \\
$\alpha$ & $10^{-2}$ & & accretion alpha disc \\
$\alpha_\mathrm{mid}$ & $10^{-4}$ & & midplane turbulence \\
$\xi$ & $0.015$ & & pebble to stellar accretion flux ratio \\
$v_\mathrm{frag}$ & $1$ & $\mathrm{m}\,\mathrm{s}^{-1}$ & fragmentation velocity \\
$\kappa$ & $0.005$ & $\mathrm{kg}^{-1}\,\mathrm{m}^2$ & envelope opacity\\
$a$ & $10{-}50$ & $\mathrm{AU}$ & initial semi-major axis \\
$e_\mathrm{rms}$ & $10^{-5}$ & & root-mean square eccentricity \\
$i_\mathrm{rms}$ & $5{\times}10^{-6}$ & $\mathrm{rad}$ & root-mean square inclination \\
$p_\mathrm{eff}$ & $0.1{-}1$ & & planetesimal formation efficiency \\
$p$ & $0.6$ & & power-law exponent IMF \\
$q$ & $0.35$ & & exponential cut-off IMF \\
$\rho_\mathrm{p}$ & $2$ & $\mathrm{g}\,\mathrm{cm}^{-3}$ & material density planetesimals and planets \\
\hline
\end{tabularx}
\end{table}

\subsection{Disc model}
\label{sec:discmodel}

Viscous disc models are widely used to describe the structure of protoplanetary discs. In the steady state, the surface density, $\Sigma_\mathrm{g}$, is determined by the accretion rate as
\begin{equation}
\Sigma_\mathrm{g}=\frac{\dot{M}_\star}{3\pi\nu},
\end{equation}
where $\nu{=}\alpha H^2\Omega_\mathrm{K}$ is the viscosity of the gas in the $\alpha$-disc model \citep{Shakura1973}. The parameter $\alpha$ is a non-dimensional measure for the accretion velocity of the disc gas and is typically ${\sim}10^{-2}$. The scale height, $H$, of the gas is the ratio of the sound speed, $c_\mathrm{s}$, and the Keplerian frequency, $\Omega_\mathrm{K}$: $H{=}c_\mathrm{s}/\Omega_\mathrm{k}$. In order to use the steady-state approximation and the derived scaling laws, it is implicitly assumed that the disc size is larger than the simulated region of interest. Outside of the characteristic radius, the disc would expand instead \citep{LyndenBell1974}.

\citet{Hartmann2016} provides a fit to stellar accretion rates with time as
\begin{equation}
\log_{10}\left(\frac{\dot{M}_\star}{M_\odot\,\mathrm{yr}^{-1}}\right)=-7.9-1.07\log_{10}\left(\frac{t}{\mathrm{Myr}}\right), \label{eq:MdotHartmann}
\end{equation}
scaled to a solar mass star. This gives $\dot{M}_\star{\approx}4.6{\times}10^{-8}\,M_\odot\,\mathrm{yr}^{-1}$ at $0.3\,\mathrm{Myr}$ and ${\sim}3.0{\times}10^{-9}\,M_\odot\,\mathrm{yr}^{-1}$ at $3.8\,\mathrm{Myr}$, which is in agreement with the lifetime of the solar nebula \citep{Wang2017}.

The scaling laws for surface density and temperature of an accretion disc in steady state are taken from \citet{Ida2016} and are based on the works of \citet{Chambers2009} and \citet{Oka2011}. In the inner disc, the temperature is set by viscous accretion, while the outer part of the disc is heated by stellar irradiation. When the accretion rate is $4.6{\times}10^{-8}\,M_\odot\,\mathrm{yr}^{-1}$, the boundary between the viscously heated and the illuminated disc will be at ${\sim}2.5\,\mathrm{AU}$. As the mass accretion rate decreases to ${\sim}3{\times}10^{-9}\,M_\odot\,\mathrm{yr}^{-1}$, the boundary shifts inwards to ${\sim}0.25\,\mathrm{AU}$ at $3.8\,\mathrm{Myr}$ (using Eq.~\ref{eq:MdotHartmann} for $\dot{M}_\star$). Because we are interested in the giant-planet formation region ${\gtrsim}10\,\mathrm{AU}$ that is well outside the viscously heated regime, we only take the irradiation into account. The power-law prescription for surface density has the form
\begin{equation}
    \Sigma_\mathrm{g}=2.7\times10^3\,\mathrm{g}\,\mathrm{cm}^{-2}\,\left(\frac{\alpha}{10^{-3}}\right)^{-1} \left(\frac{\dot{M}_\star}{10^{-8}\,M_\odot\,\mathrm{yr}^{-1}}\right) \left(\frac{r}{\mathrm{AU}}\right)^{-15/14}.
\end{equation}
Here, the parameter $\alpha$ is the non-dimensional measure for the accretion velocity of the disc gas. The mass accretion rate is given by Eq.~\ref{eq:MdotHartmann}. The irradiated disc temperature \citep{Ida2016} takes the form
\begin{equation}
    T=150\,\mathrm{K}\,\left(\frac{r}{\mathrm{AU}}\right)^{-3/7}.
\end{equation}
The pressure support of the disc is expressed as
\begin{equation}
    \eta=-\frac{1}{2}\left(\frac{H}{r}\right)^2\frac{\partial \log P}{\partial \log r}, \label{eq:gaseta}
\end{equation}
with $P$ being the gas pressure and $H{/}r$ the aspect ratio of the gas disc. The logarithmic pressure gradient simplifies to $\partial \log P/\partial \log r{\approx}{-}2.8$ for the given exponents of the surface density and temperature profiles.

\subsection{Pebble accretion}
\label{sec:pebbleaccretion}

The pebble accretion prescription is based on the work of \citet{Lambrechts2019}. We distinguish between the 3D accretion regime for low-mass embryos and the 2D regime for higher mass embryos. The first step to calculate effective pebble accretion rates is to determine pebbles sizes throughout the disc, which can differ significantly across the full disc range considered in this work.

\subsubsection{Pebble Stokes number}
\label{sec:pebblestokesnumber}

The maximum size of the pebbles is determined by how fast micron-sized grains grow, fragment, and drift radially. Following the works of \citet{Birnstiel2012}, \citet{Lambrechts2014b}, and \citet{Drazkowska2021}, we assume that initially all grains are (sub-)micron in size with Stokes number $\mathrm{St}_0$ and grow exponentially according to
\begin{equation}
    \mathrm{St}_\mathrm{growth}=\mathrm{St}_0\exp\left(\frac{t}{t_\mathrm{growth}}\right),
\end{equation}
with a growth timescale, $t_\mathrm{growth}$, of
\begin{equation}
    t_\mathrm{growth}=\frac{\Sigma_{\mathrm{g},0}}{\Sigma_{\mathrm{d},0}}\Omega_\mathrm{K}^{-1}\left(\frac{\alpha_\mathrm{mid}}{10^{-4}}\right)^{-1/3}\left(\frac{r}{\mathrm{AU}}\right)^{1/3}.
\end{equation}

As the dust grows, collision velocities increase, and the maximum Stokes number is reached when the turbulent collision speeds equal the fragmentation velocity of the pebbles:
\begin{equation}
    \mathrm{St}_\mathrm{frag}=0.37\frac{v_\mathrm{frag}^2}{3\alpha_\mathrm{mid}c_\mathrm{s}^2}, \label{eq:stfrag}
\end{equation}
where $v_\mathrm{frag}$ is the fragmentation velocity, which we assumed to be $1\,\mathrm{m}\,\mathrm{s}^{-1}$. Laboratory experiments argue for ${\sim}1\,\mathrm{m}\,\mathrm{s}^{-1}$ for the fragmentation velocity of silicates \citep{Blum2008}. For icy pebbles, however, the value of the fragmentation threshold velocity is debated and values range from ${\sim}10\,\mathrm{m}\,\mathrm{s}^{-1}$ \citep{Gundlach2015} to as high as ${\sim}50\,\mathrm{m}\,\mathrm{s}^{-1}$ \citep{Wada2009}. However, more recent studies have shown that the sticking properties of water ice may be comparable to that of silicates for cold disc conditions, as is the case in the giant-planet region \citep{Gundlach2018,Musiolik2019,Kimura2020}. We therefore use $1\,\mathrm{m}\,\mathrm{s}^{-1}$ in our study.

If pebble drift faster than they can grow, the limiting Stokes number is
\begin{equation}
    \mathrm{St}_\mathrm{drift}=\frac{1}{30\eta}\frac{\Sigma_\mathrm{d}}{\Sigma_\mathrm{g}}.
\end{equation}
The factor $1/30$ comes from the fact that the growth timescale needs to be ${\sim}30$ times faster than the drift timescale for pebble growth to outperform radial drift \citep{Okuzumi2012}. When taking all the limiting cases into account, the actual Stokes number of the pebbles is
\begin{equation}
    \mathrm{St}=\max\left(\mathrm{St}_0,\min\left(\mathrm{St}_\mathrm{growth},\mathrm{St}_\mathrm{frag},\mathrm{St}_\mathrm{drift}\right)\right),
\end{equation}
where we include $\mathrm{St}_0$ because pebbles cannot be smaller than a single grain.

\subsubsection{Pebble accretion rates}

In the 3D regime, we use the minimum of the 3D accretion rate from \citet[their Eq.~A.13]{Lambrechts2019}, $\dot{m}_\mathrm{L19}$, and the accretion rate for the accretion radius in the weak-coupling Bondi regime, $\dot{m}_\mathrm{wc}$, found from their Eq.~A.5:
\begin{align}
    \dot{m}_\mathrm{L19}&=\frac{1}{2\sqrt{2\pi}}\frac{m}{M_\odot}\left(\frac{H_\mathrm{peb}}{H}\right)^{-1}\left(\frac{H}{r}\right)^{-1}\frac{1}{\eta}F_\mathrm{peb}, \label{eq:mdotl19} \\
    \dot{m}_\mathrm{wc}&=\pi\Sigma_\mathrm{p}r_\mathrm{acc,wc}^2\left(v_\mathrm{rel}+\frac{3}{2}\Omega_\mathrm{K}r_\mathrm{acc,wc}\right)\frac{\Sigma_\mathrm{peb}}{\sqrt{2\pi}H_\mathrm{peb}}.
\end{align}
Here, $H_\mathrm{peb}/H$ is the ratio of pebble scale-height and gas scale-height, $\eta$ is related to the pressure gradient in the disc (Eq.~\ref{eq:gaseta}), and $F_\mathrm{\mathrm{peb}}$ is the pebble flux.

The transition mass above which accretion changes from 3D to 2D to distinguish between the two regimes is
\begin{equation}
    m_\mathrm{2D}=\frac{\left(\sqrt{\tfrac{8}{\pi}}H_\mathrm{peb}\right)^2\left(\delta v_\mathrm{p}+\frac{3}{2}\Omega_\mathrm{K}\sqrt{\tfrac{8}{\pi}}H_\mathrm{peb}\right)}{G t_\mathrm{f}},
\end{equation}
where $t_\mathrm{f}{=}\mathrm{St}\Omega_\mathrm{K}^{-1}$ is the friction time of the pebbles, the accretion radius is $r_\mathrm{acc}{=}\sqrt{8/\pi}H_\mathrm{peb}$ with pebble scale-height $H_\mathrm{peb}$, and the accretion velocity is the velocity of the planet relative to the local circular orbit, $\delta v_\mathrm{p}$, plus the shear at the accretion radius. For a circular, planar orbit $\delta v_\mathrm{p}{=}0$.

In the 2D regime, we used the maximum of the headwind-dominated, $\dot{m}_\mathrm{hw}$, and shear-dominated, $\dot{m}_\mathrm{sh}$, accretion rates given as
\begin{align}
    \dot{m}_\mathrm{hw}&=2\Sigma_\mathrm{peb}\sqrt{2Gmt_\mathrm{f}v_\mathrm{rel}}, \\
    \dot{m}_\mathrm{sh}&=2\Sigma_\mathrm{peb}\frac{3}{2}\Omega_\mathrm{K}\left(\frac{4Gmt_\mathrm{f}}{3\Omega_\mathrm{K}}\right)^{2/3}, \label{eq:mdotshear}
\end{align}
where $v_\mathrm{rel}$ is the relative velocity between the planet and the gas and $m$ is the mass of the planet.
Summarising Eqs.~\ref{eq:mdotl19} to \ref{eq:mdotshear} for the 3D and 2D regimes, the actual pebble accretion rate is
\begin{align}
    \dot{m}=
    \begin{cases}
        \min\left(\dot{m}_\mathrm{L19},\dot{m}_\mathrm{wc}\right) & \left(m\le m_{2\mathrm{D}}\right) \\
        \max\left(\dot{m}_\mathrm{hw},\dot{m}_\mathrm{sh}\right) & \left(m> m_{2\mathrm{D}}\right)
    \end{cases}
    .
\end{align}

\subsubsection{Pebble isolation mass}

Pebble accretion stops when the planet reaches the pebble isolation mass, $M_\mathrm{iso}$ \citep{Lambrechts2014a}. \citet{Bitsch2018} finds 
\begin{align}
    m_\mathrm{iso}&=25M_\oplus\left(\frac{H/r}{0.05}\right)^3 \nonumber \\
    &\times\left(0.34\left(\frac{-3}{\log_{10}\alpha_\mathrm{mid}}\right)^4+0.66\right)\left(1-\frac{\tfrac{\partial\ln{P}}{\partial\ln{r}}+2.5}{6}\right) \label{eq:pebisomass}.
\end{align}
Here, $P$ is the gas pressure and $\alpha_\mathrm{mid}$ is the turbulent stirring strength of pebbles in the mid-plane.

\subsubsection{Reduction of pebble flux}

We include the reduction of the pebble flux through accreting planets as follows. The outermost planet receives the full pebble flux $F_1{=}F_{\mathrm{peb}}$ of which it accretes per unit of time a fraction $x_1$ such that $\dot{m}_1{=}x_1F_1$. This reduces the pebble flux for the second outermost planet to $F_2{=}F_1{-}\dot{m}_1{=}F_1(1{-}x_1)$. This planet would then accrete $\dot{m}_2{=}x_2F_2{=}x_2(1{-}x_1)F_1$. Extending this procedure, the pebble flux planet $i$ receives is
\begin{equation}
    F_i=F_\mathrm{peb}\prod_{k=1}^{i-1}(1-x_k). \label{eq:reducedpebflux}
\end{equation}
A planet that reaches pebble isolation mass cuts off the pebble flux to the planets located inside of its orbit. This approximation simplifies the interpretation of the results. However, coagulation/fragmentation studies show that small dust fragments may cross the planetary gap \citep{Stammler2023}.

\subsubsection{Pebble flux}

\begin{figure}
\resizebox{\hsize}{!}{\includegraphics{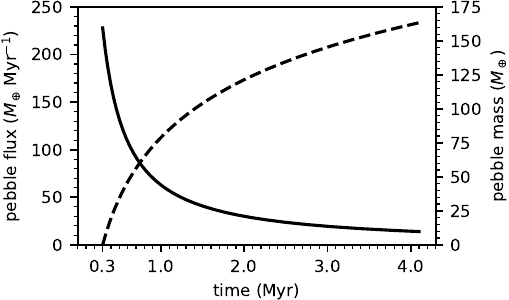}}
\caption{Pebble flux (solid) and total pebble mass (dashed) drifting through the system over time.
}
\label{fig:pebflux}
\end{figure}
In our model, the pebble flux is set to a fixed fraction of the stellar mass accretion rate,
\begin{equation}
    F_\mathrm{peb}=\xi\dot{M}_\star, \label{eq:pebbleflux}
\end{equation}
with $\dot{M}_\star$ given by Eq.~\ref{eq:MdotHartmann} and $\xi{=}1.5\%$. Using viscous disc theory, it can be shown that the flux ratio $\xi$ is related to the ratio of pebble and gas surface density:
\begin{equation}
    \frac{\Sigma_\mathrm{peb}}{\Sigma_\mathrm{g}}=\frac{\xi}{\left(2/3\right)|\partial\ln P/\partial\ln r|\left(\mathrm{St}/\alpha\right)+1}
\end{equation}
\citep{Johansen2019a}. Here, $\alpha$ is the accretion disc $\alpha$ value and not the midplane turbulence. In the case where $\mathrm{St}{\lesssim}\alpha$, the flux ratio represents the metallicity $\Sigma_\mathrm{peb}/\Sigma_\mathrm{g}$ of the disc, which is in our case $1.5\%$, slightly higher than the standard value of $1\%$. For $\mathrm{St}{\gtrsim}\alpha$, the local metallicity drops below the original value. Alternatively, the pebble flux could be set by fixing a pebble production rate at the outer edge of the disc \citep{Lambrechts2014b,Ida2016}. In Fig.~\ref{fig:pebflux} we show the temporal evolution of the pebble flux and the total pebble mass drifting through the system. The pebble flux decreases from initially ${\sim}230\,M_\oplus\,\mathrm{Myr}^{-1}$ to ${\sim}14\,M_\oplus\,\mathrm{Myr}^{-1}$ at the end of the simulation. By integrating Eq.~\ref{eq:pebbleflux}, the total mass of pebbles drifting through the system is ${\sim}163\,M_\oplus$.

\subsection{Migration}
\label{sec:migration}

Migration, eccentricity, and inclination damping are implemented by applying additional accelerations,
\begin{align}
    \mathbf{a}_\mathrm{m}&=-\frac{\mathbf{v}}{t_\mathrm{m}}, \\
    \mathbf{a}_e&=-2\frac{\left(\mathbf{v}\cdot\mathbf{r}\right)\mathbf{r}}{r^2t_e}, \\
    \mathbf{a}_i&=-\frac{v_z}{t_i}\mathbf{k},
\end{align}
to each body. Here, $t_\mathrm{m}$, $t_e$, and $t_i$ are the respective timescales for migration and damping, and $\mathbf{k}$ is the unit vector in $z$ direction \citep{Cresswell2008}.

\paragraph{Type I}

We used the timescales for migration, eccentricity, and inclination damping from \citet{Cresswell2008}. The general damping timescale due to planet-disc interaction is
\begin{equation}
    t_\mathrm{wave}=\frac{M_\star}{m}\frac{M_\star}{\Sigma r^2}\left(\frac{H}{r}\right)^4\Omega_\mathrm{K}^{-1}
\end{equation}
\citep{Tanaka2004}. The damping timescale for eccentricity and inclination are
\begin{align}
    t_{e}&=\frac{t_\mathrm{wave}}{0.780} \nonumber \\
    &\times\left[1{-}0.14\left(\frac{e}{H/r}\right)^2{+}0.06\left(\frac{e}{H/r}\right)^3{+}0.18\left(\frac{e}{H/r}\right)\left(\frac{i}{H/r}\right)^2\right], \label{eq:tdampe} \\
    t_{i}&=\frac{t_\mathrm{wave}}{0.544}\nonumber \\
    &\times\left[1{-}0.30\left(\frac{i}{H/r}\right)^2{+}0.24\left(\frac{i}{H/r}\right)^3{+}0.14\left(\frac{e}{H/r}\right)^2\left(\frac{i}{H/r}\right)\right], \label{eq:tdampi}
\end{align}
and the migration timescale is
\begin{align}
    t_\mathrm{m}&=\frac{2t_\mathrm{wave}}{k_\mathrm{m}}\left(\frac{H}{r}\right)^{-2}\left\{P(e)+\frac{P(e)}{|P(e)|}\right.\\
    &\times\left.\left[0.070\left(\frac{i}{H/r}\right){+}0.085\left(\frac{i}{H/r}\right)^4{-}0.080\left(\frac{e}{H/r}\right)\left(\frac{i}{H/r}\right)^2\right]\right\}, \nonumber
\end{align}
where
\begin{equation}
    P(e)=\frac{1+\left(\frac{e}{2.25H/r}\right)^{1.2}+\left(\frac{e}{2.84H/r}\right)^6}{1-\left(\frac{e}{2.02H/r}\right)^4}
\end{equation}
\citep{Cresswell2008} and $k_\mathrm{m}{=}1.36{+}0.62q_\mathrm{S}{+}0.43p_\mathrm{T}$ \citep{DAngelo2010}, with $q_\mathrm{S}{=}{15/14}$ and $p_\mathrm{T}{=}{3/7}$ being the power-law exponents of surface density and temperature, respectively.

\paragraph{Type II}

The transition to type-II migration occurs for (partially) gap-opening planets exceeding the pebble isolation mass. We implement this transition by multiplying the type-I migration timescale, $t_\mathrm{m}$, with the correction factor
\begin{equation}
    \frac{\Sigma_\mathrm{gap}}{\Sigma}=\frac{1}{1+\left(m/(2.3\,m_\mathrm{iso})\right)^2},
\end{equation}
which is the ratio of the gas surface density in the gap, $\Sigma_\mathrm{gap}$, to the unperturbed surface density \citep{Kanagawa2018,Ida2018,Johansen2019a}.

\subsection{Gas accretion}
\label{sec:gasaccretion}

The planets in our simulations were allowed to accrete gas once they reached pebble isolation mass. In the first phase, the atmosphere grows through Kelvin-Helmholtz contraction:
\begin{equation}
    \dot{m}_\mathrm{KH}=10^{-5}\,M_\oplus\,\mathrm{yr}^{-1}\left(\frac{m}{100\,M_\oplus}\right)^4\left(\frac{\kappa}{0.1\,\mathrm{kg}^{-1}\,\mathrm{m}^2}\right)^{-1}
\end{equation}
\citep{Ikoma2000}. Here, $\kappa$ is the opacity of the atmosphere.
 
A second contribution to gas accretion is the accretion of the disc gas that enters the Hill sphere of the planet during runaway gas accretion. \citet{Lambrechts2019b} provide a fit to hydrodynamic simulations that quantifies the amount of the gas that is bound in the growing atmosphere:
\begin{equation}
    \dot{m}_{\mathrm{H}}=2.7\times10^{-3}\,M_\oplus\,\mathrm{yr}^{-1}\left(\frac{m}{100\,M_\oplus}\right)^{1.9}.
\end{equation}

Lastly, the maximum amount of gas that can be accreted is provided by the stellar gas accretion rate, $\dot{M}_\star$. We therefore used the minimum of the three accretion rates as the actual gas accretion rate of the planet:
\begin{equation}
    \dot{m}=\min(\dot{m}_\mathrm{KH},\dot{m}_{\mathrm{H}},\dot{M}_\star).
\end{equation}

\subsection{Planetesimal sizes}
\label{sec:planetesimalsizes}

Collisional growth of dust and ice grains forms millimetre- to centimetre-sized pebbles throughout the disc. These pebbles are concentrated in filaments by the streaming instability and subsequently collapse gravitationally into planetesimals \citep{Youdin2005,Johansen2014,Li2018}. The initial mass function (IMF) of planetesimals is an ongoing research question \citep{Johansen2015,Schaefer2017,Li2019,Simon2022,Gerbig2023}. While the steep upper end of the IMF is found to be similar to the observed size distribution of cold classical Kuiper belt objects \citep{Kavelaars2021}, the lower end depends on resolution \citep{Li2019}. Here we use an exponentially tapered power law \citep{Schaefer2017}. For masses below a characteristic planetesimal mass, the IMF follows a power law with exponent ${\sim}0.6$. Above the characteristic mass, the IMF drops exponentially limiting the maximum mass to ${\sim}10^{-3}{-}10^{-1}\,M_\oplus$, depending on the distance from the star. A note on nomenclature, here we use the term embryo for the top of the IMF, bodies usually in the range $10^{-3}{-}10^{-1}\,M_\oplus$.

The IMF can be expressed as the cumulative number of bodies with a mass higher than $m$:
\begin{equation}
    \frac{N(>m)}{N_\mathrm{tot}}=\left(\frac{m}{m_\mathrm{min}}\right)^{-p}\exp\left[\left(\frac{m_\mathrm{min}}{m_\mathrm{p}}\right)^q-\left(\frac{m}{m_\mathrm{p}}\right)^q\right], \label{eq:imf}
\end{equation}
where $m_\mathrm{p}$ is the characteristic planetesimal mass at which the exponential drop-off starts and $m_\mathrm{min}{\ll}m_\mathrm{p}$ is a minimum mass \citep{Schaefer2017}.
We used an approximate scaling law for the characteristic planetesimal mass, $m_\mathrm{p}$, that was derived in \citet{Liu2020} by compiling streaming instability simulations. The expression for $m_\mathrm{p}$ reads as
\begin{equation}
    m_\mathrm{p}=5{\times}10^{-5}\,M_\oplus\left(\frac{C}{5{\times}10^{-5}}\right)\left(\frac{Z}{0.02}\right)^{0.5}\left(\frac{\gamma}{\pi^{-1}}\right)^{1.5}\left(\frac{H/r}{0.05}\right)^3\left(\frac{M_\star}{M_\odot}\right),
\end{equation}
where $C$ is a parameter to match the scaling law to streaming instability simulations, $Z$ is the metallicity of a streaming instability filament, $\gamma{=}4\pi G\rho_\mathrm{g}/\Omega_\mathrm{K}^2$ is a measure for the relative strength between self-gravity and shear and $M_\star$ is the stellar mass. To convert from mass to radius, we use a constant density of $2\,\mathrm{g}\,\mathrm{cm}^{-3}$. We do not employ an explicit mass-radius relationship \citep[e.g.][]{Seager2007}.

\subsection{GPU-accelerated \texorpdfstring{$N$}{N}-body simulations}
\label{sec:gpunbodysimulations}

We used the Gravitational Encounters with GPU Acceleration (\texttt{GENGA}) code \citep{Grimm2014,Grimm2022} for $N$-body simulations of planet formation. We added the effects of pebble accretion, gas accretion, and migration to simulate the formation process in the nebular phase. Using \texttt{GENGA}, we were able to include several thousand bodies in the simulations to self-consistently follow the growth of a size distribution of bodies, drawn from streaming instability simulations as described in detail in Sect.\,\ref{sec:planetesimalsizes}, ranging from ${\sim}200\,\mathrm{km}$-sized planetesimals to ${\sim}0.1\,M_\oplus$ embryos.

We implemented our pebble and gas accretion prescriptions as \texttt{CUDA} kernels in \texttt{GENGA}. The new kernels are called each time step in the $N$-body step of \texttt{GENGA}. While the gas accretion kernel affects each planet individually, the reduction of the pebble flux in the pebble accretion kernel requires knowledge of the radial order of all the planets in the simulation. To achieve this, we firstly calculated the accretion efficiency $x_i{=}\dot{m}_i/F_i$ of each pebble-accreting planet, which is independent of the pebble flux. In a next step, we sorted the planets according to their radial distance from the central star. With the sorted list of planets, we proceeded by calculating the reduced pebble flux for each planet (Eq.~\ref{eq:reducedpebflux}). Finally, we calculated the accreted mass in pebbles by multiplying the accretion efficiency $x_i$ with the reduced flux $F_i$.

\subsection{Initial conditions}
\label{sec:initialconditions}

We placed the simulations in the giant-planet formation region between $10$ and $50\,\mathrm{AU}$ and integrated the nebular phase for $3.8\,\mathrm{Myr}$ from $0.3\,\mathrm{Myr}$ until $4.1\,\mathrm{Myr}$ with a time step of $12$ days. This assumed lifetime of the gas disc lies in the estimated range of lifetimes for the gas disc around the Sun of $2.51{-}4.89\,\mathrm{Myr}$ after CAI formation (i.\,e.\,after $t{=}0$; \citealt{Wang2017,Weiss2021}). After $4.1\,\mathrm{Myr}$, the runs were extended to $100\,\mathrm{Myr}$ in the gas-free phase.

We explored two initial spatial distributions for the planetesimals. One setup where the planetesimals were distributed in spatially separated narrow rings (narrow rings henceforth) and one where we started with a uniform distribution in a single wide ring (wide ring henceforth). For narrow rings, we placed all planetesimals in four rings each of width $\eta r$ that is the typical width of a streaming instability filament \citep{Yang2014,Liu2019,Gerbig2020}. For the wide rings, we placed the planetesimals such that the rings connected. Each ring had constant surface density determined by the filament metallicity $Z_\mathrm{fil}$ meaning that the initial surface density of the planetesimals followed the gas surface density with slope $15/14({\approx}1)$.

\begin{figure}
\resizebox{\hsize}{!}{\includegraphics{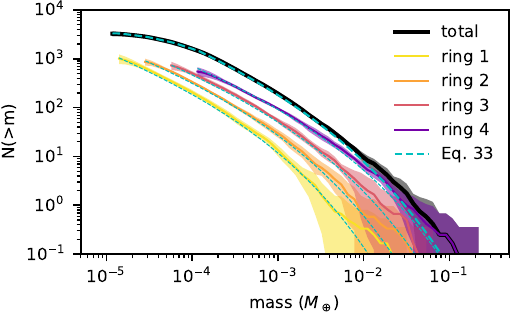}}
\caption{Initial planetesimal mass distribution. We show the initial distributions of the rings (solid coloured lines), the total distribution of all bodies (black), and Eq.~\ref{eq:imf} (dashed cyan lines). All distributions have been averaged over all low-mass runs. The standard deviation (shaded area) shows the spread of the initial planetesimal mass distribution from the random sampling. The high-mass runs follow the same IMF but with a higher total number of bodies (see also Fig.~\ref{fig:finalccdf}).
}
\label{fig:initialccdf}
\end{figure}

In each case, we numerically sampled the IMF, where we set the minimum mass $m_\mathrm{min}{=}5{\times}10^{-3}\,m_\mathrm{p}$, $p{=}0.6$, and $q{=}0.35$ \citep{Schaefer2017}. This led to planetesimal populations ranging from ${\sim}200\,\mathrm{km}$ to about $0.1\,M_\oplus$, depending on semi-major axis, with small planetesimals dominating in number, but larger embryos dominating in mass. Figure~\ref{fig:initialccdf} shows the initial complementary cumulative mass distribution (CCDF) of planetesimals averaged over all runs in the wide and narrow ring simulations. The maximum embryo mass is of the order ${\sim}0.1\,M_\oplus$ (${\sim}$Mars mass) while $50\%$ of the mass is in planetesimals less massive than ${\sim}2{\times}10^{-3}\,M_\oplus$ (${\sim}$Pluto mass). The maximum embryo mass varied from run to run because of the random sampling of the IMF.

To determine the total mass of the planetesimals, we first determined the total mass of a streaming instability filament: 
\begin{equation}
    M_\mathrm{fil}=2\pi r\Sigma_\mathrm{g}Z_\mathrm{fil}\eta r.
\end{equation}
Here we assumed that the dust-to-gas ratio of the filament, $Z_\mathrm{fil}$, is given as the maximum of, firstly, the Stokes-number-dependent criterion of \citet{Yang2017},
\begin{equation}
    \log{Z_\mathrm{c}}=
    \begin{cases}
        0.30\left(\log\mathrm{St}\right)^2+0.59\log\mathrm{St}-1.57 & \left(\mathrm{St}{>}0.1\right), \\
        0.10\left(\log\mathrm{St}\right)^2+0.20\log\mathrm{St}-1.76 & \left(\mathrm{St}{<}0.1\right),
    \end{cases}
\end{equation}
and, secondly, the dust-to-gas ratio that is obtained from balancing vertical settling and diffusion, $Z_\mathrm{sett}{=}\sqrt{\alpha_\mathrm{mid}/\left(\alpha_\mathrm{mid}+\mathrm{St}\right)}$.

In this way, we obtained $Z_\mathrm{fil}{=}\max\left(Z_\mathrm{c},Z_\mathrm{sett}\right)$. The total mass in planetesimals was then $M_\mathrm{tot}{=}p_\mathrm{eff}\,M_\mathrm{fil}$, with $p_\mathrm{eff}$ describing a planetesimal formation efficiency that is not very well constrained. Variations from ${\lesssim10}\,\%$ to $80\,\%$ have been observed in simulations \citep{Abod2019}. With $p_\mathrm{eff}{=}10\,\%$, representing the lower end of possible values, we end up with a total mass of planetesimals in the $10{-}50\,\mathrm{AU}$ region of ${\sim}1\,M_\oplus$. With $p_\mathrm{eff}{=}100\,\%$, we would obtain $M_\mathrm{tot}{\sim}10\,M_\oplus$, which we used for the high-mass simulations.

Eccentricities and inclinations were sampled from Rayleigh distributions with respective root mean square values of $10^{-5}$ and $5{\times}10^{-6}$. We chose low initial values of eccentricity and inclination because the combination of mutual scattering and gas drag leads to rapid excitation and damping, rendering the initial choice unimportant. Arguments of perihelion, longitudes of ascending node, and mean anomalies were sampled from uniform distributions in the range from $0$ to $2\pi$.

For each initial configuration, narrow rings and wide rings, we ran a total of six simulations with randomised initial configurations of planetesimals. This provides a small statistical sample for further analysis. The simulations are labelled \texttt{n05} to \texttt{n10} for narrow ring runs and \texttt{w03} to \texttt{w08} for wide ring runs. We ran an additional set of six simulations with a total planetesimal mass of ${\sim}10\,M_\oplus$, which are labelled \texttt{m01} to \texttt{m06}. The low-mass simulations typically contained around $4000$ fully interacting bodies. The high-mass simulations were divided into ${\sim}4000$ fully interacting bodies and ${\sim}36\,000$ semi-active test particles that did not interact with each other but interacted gravitationally with the fully interacting bodies. The mass below which bodies are considered test particles was set to $10^{-3}\,M_\oplus$, which is approximately the transition mass below which pebble accretion becomes inefficient evaluated at ${\sim}3.6\,\mathrm{AU}$ \citep{Lambrechts2012,Ormel2017}.

The high number of bodies per simulation led to run times of several days for the low-mass rings and up to weeks for the high-mass rings, which rendered it difficult to obtain larger sets due to limited computational resources. We used Nvidia A30 GPUs in Multi-Instance GPU (MIG) configuration of four instances for the low-mass runs. The run times ranged from $33$ to $131$ hours ($1.4{-}5.5$ days) with an average of $56.2$ hours ($2.3$ days) per simulation. For the high-mass runs we used the full capacity of the Nvidia A30 GPU. The run times were $146$ to $309$ hours ($6.1{-}12.9$ days) with an average of $197$ hours ($8.2$ days) per simulation. The total number of $115\,662\,000$ time steps and the output interval of one output every $100\,000$ time steps were the same for each low- and high-mass run. Table~\ref{tab:initialconditions} summarises the parameters used in each simulation and Table~\ref{tab:initialrings} lists the initial ring locations for the narrow and wide ring runs.
\begin{table}
\caption{Initial ring locations.}
\label{tab:initialrings}
\centering
\begin{tabularx}{\columnwidth}{llllll}
\hline\hline
run & ring 1 & ring 2 & ring 3 & ring 4 & $M_\mathrm{tot}$\\
& $(\mathrm{AU})$ & $(\mathrm{AU})$ & $(\mathrm{AU})$ & $(\mathrm{AU})$ & ($M_\oplus$) \\
\hline
\texttt{n05}{-}\texttt{n10} & $12.2$ & $18.3$ & $27.3$ & $40.9$ & $1$ \\
\texttt{w03}{-}\texttt{w08} & $10{-}15$ & $15{-}22$ & $22.4{-}33$ & $33{-}50$ & $1$ \\
\texttt{m01}{-}\texttt{m06} & $10{-}15$ & $15{-}22$ & $22.4{-}33$ & $33{-}50$ & $10$ \\
\hline
\end{tabularx}
\end{table}

\section{Results}
\label{sec:results}

To describe the planets formed in our simulations, we used mass categories to cover planetesimals (${<}10^{-3}\,M_\oplus$), Pluto- to Mars-mass bodies ($10^{-3}{-}0.1\,M_\oplus$), terrestrial-like planets ($0.1{-}2\,M_\oplus$), super-Earths ($2{-}10\,M_\oplus$), ice-giants ($10{-}80\,M_\oplus$), and gas giants (${>}80\,M_\oplus$). This was motivated by the different types of planets found in the Solar System and exoplanetary systems. Because $N$-body simulations are stochastic, the outcome of individual simulations varies in terms of, for example, the number of planets formed or their final configuration. Before summarising and comparing the outcome of each set of simulations, we describe the typical evolution for the case of a single run.

\subsection{Example narrow ring run \texttt{n05}}

\subsubsection{Nebular phase}
\begin{figure*}
\centering
\includegraphics[width=17cm]{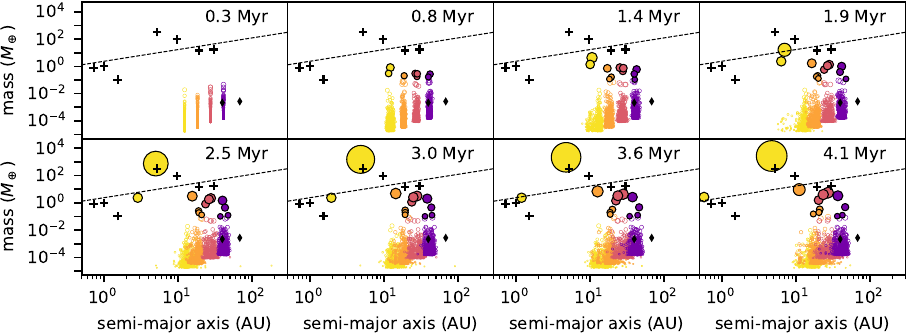}
\caption{
Snapshots of mass and semi-major axis from $0.3\,\mathrm{Myr}$ to $4.1\,\mathrm{Myr}$ for narrow ring run \texttt{n05}. The initial locations of the rings are $12.2\,\mathrm{AU}$ (yellow), $18.3\,\mathrm{AU}$ (orange), $27.3\,\mathrm{AU}$ (red), and $40.9\,\mathrm{AU}$ (purple). The symbol size scales with the mass of the body as ${\propto}m^{1/3}$. Planets with mass ${\ge}0.1\,M_\oplus$ have solid circles. The Solar System planets ($+$) and Pluto and Eris ($\Diamond$) are shown for reference as well as the pebble isolation mass (dashed line).
}
\label{fig:snapnarrowm}
\end{figure*}
\begin{figure*}
\centering
\includegraphics[width=17cm]{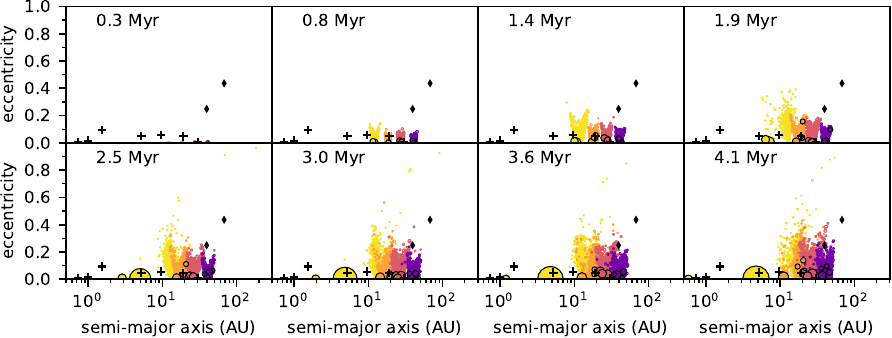}
\caption{
Same as Fig.~\ref{fig:snapnarrowm} but showing the eccentricity and semi-major axis evolution.
}
\label{fig:snapnarrow}
\end{figure*}
In the nebular phase, gas is present until disc dissipation at $4.1\,\mathrm{Myr}$. Figure~\ref{fig:snapnarrowm} illustrates core growth due to pebble accretion and mutual collisions in narrow ring run \texttt{n05}. Figure~\ref{fig:snapnarrow} shows the eccentricity versus semi-major axis for all bodies in this simulation.

From initially narrow rings with low eccentricity, the top of the mass distribution grows by collisions and pebble accretion. These embryos, while embedded in the rings, start to dynamically excite the smaller planetesimals to higher eccentricities (and inclinations). While the lower-mass planetesimals gain higher eccentricities and inclinations, the higher-mass embryos acquire lower eccentricities and inclinations because of dynamical friction and eccentricity and inclination damping \citep{Ida1992b,Cresswell2008}. Furthermore, this excitation causes a significant broadening of the initially narrow rings \citep{Ohtsuki2003,Tanaka2003}, especially towards later times (${\gtrsim}1.5\,\mathrm{Myr}$).

We observe that bodies in the innermost ring located at ${\sim}12\,\mathrm{AU}$ grow faster than the ones located farther out, even though they were seeded with initially higher masses. This is a consequence of the inside-out growth behaviour of pebble accretion in an irradiated protoplanetary disc, where the pebble accretion timescale decreases with increasing distance from the star \citep{Lambrechts2014b}. As the two most massive bodies in the innermost ring grow, they start to migrate at ${\sim}1.5\,\mathrm{Myr}$, which causes depletion of the ring and smaller planetesimals to be dragged along. At ${\sim}1.9\,\mathrm{Myr}$, one of the two planets reaches pebble isolation mass marking the onset of gas accretion. This planet continues to grow and migrate until the transition to type-II migration and disc dissipation slow down and finally stop its inwards motion at ${\sim}5\,\mathrm{AU}$. The lower-mass companion also continues to migrate. Because this planet is inside the orbit of the other one, it does not continue to accrete pebbles because in our model a planet that reaches pebble isolation mass cuts off the pebble flux. Hence, this ${\sim}1{-}2\,M_\oplus$ planet does not grow and continues fast type-I migration until disc dissipation stops migration at ${\sim}0.6\,\mathrm{AU}$. Towards the very end of the simulation at ${\sim}3.7\,\mathrm{Myr}$, this planet crosses the pebble isolation mass line and accretes a tiny amount of gas.

In addition to the gas giants, several planets with masses in the terrestrial to the ice giant range form in the outer rings. These planets do not migrate far and remain closer to their birth location in the $10{-}50\,\mathrm{AU}$ region. The planet masses tend to decrease with distance, which is again a consequence of the inside-out growth. Towards the end of the nebular phase, the gas giant and the planets still embedded in their birth environment excite the remaining planetesimals and produces a scattered disc-like structure containing planetesimals from the two innermost rings, that is the population originally located between ${\sim}10\,\mathrm{AU}$ and $25\,\mathrm{AU}$.

\subsubsection{Gas-free phase}

\begin{figure*}
\centering
\includegraphics[width=17cm]{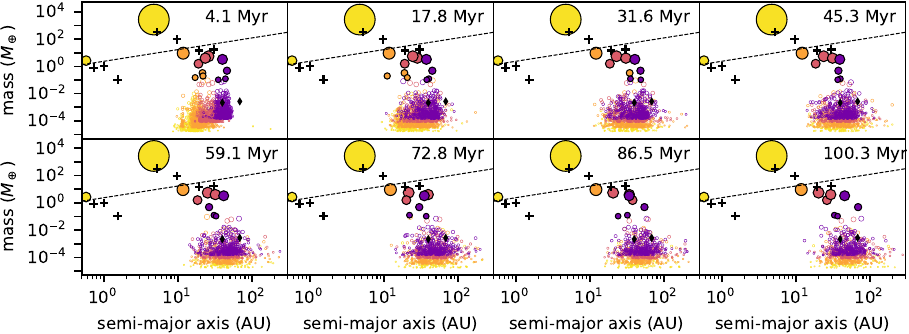}
\caption{
Same as Fig.~\ref{fig:snapnarrowm} but for $4.1\,\mathrm{Myr}$ to $100\,\mathrm{Myr}$.
}
\label{fig:snapnarrowmgf}
\end{figure*}
\begin{figure*}
\centering
\includegraphics[width=17cm]{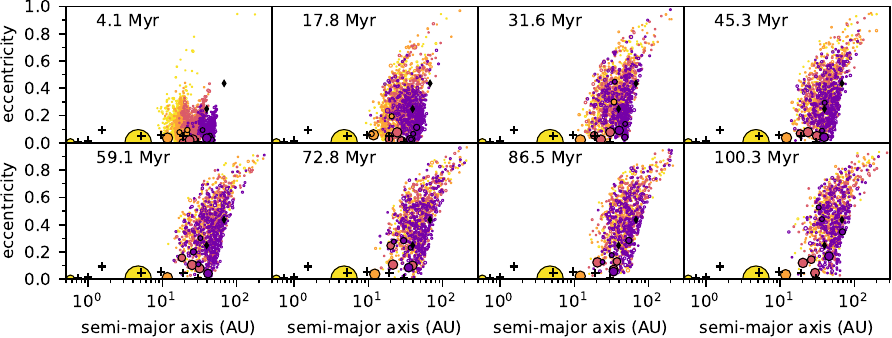}
\caption{
Same as Fig.~\ref{fig:snapnarrow} but for $4.1\,\mathrm{Myr}$ to $100\,\mathrm{Myr}$.
}
\label{fig:snapnarrowgf}
\end{figure*}
After disc dissipation at $4.1\,\mathrm{Myr}$, we integrated run \texttt{n05} in the gas-free phase until $100\,\mathrm{Myr}$. Figures~\ref{fig:snapnarrowmgf} and \ref{fig:snapnarrowgf} illustrate the evolution of mass and eccentricity versus semi-major axis for all bodies in this simulation.

When the gas disc dissipates, the damping effect of the gas on the orbits of the planets and planetesimals disappears. This leads to strong dynamical excitation of eccentricities and inclinations, depletion of planetesimals and planets by ejection, and mixing of the rings such that memory of the initial distribution is completely lost. A total mass of ${\sim}1.5\,M_\oplus$ is ejected, whereof $0.2\,M_\oplus$ are planetesimals with mass ${\le}10^{-3}\,M_\oplus$, $0.6\,M_\oplus$ are Pluto- to Mars mass bodies ($10^{-3}{-}0.1\,M_\oplus$), and $0.7\,M_\oplus$ are terrestrial-like planets ($0.1{-}2\,M_\oplus$). In numbers, ${\sim}1400$ planetesimals, $88$ Pluto-to Mars mass bodies, and $3$ terrestrial-like planets are lost over the $100\,\mathrm{Myr}$ of evolution. Noteworthy is the formation of a scattered-disc-like structure consisting of planetesimals and embedded planets extending from ${\sim}20{-}30\,\mathrm{AU}$ to up to ${\sim}200\,\mathrm{AU}$ in semi-major axis. Such a scattered disc is formed in all our simulations, regardless of the initial distribution of planetesimals.

\subsection{Collisions}

\begin{figure}
\resizebox{\hsize}{!}{\includegraphics{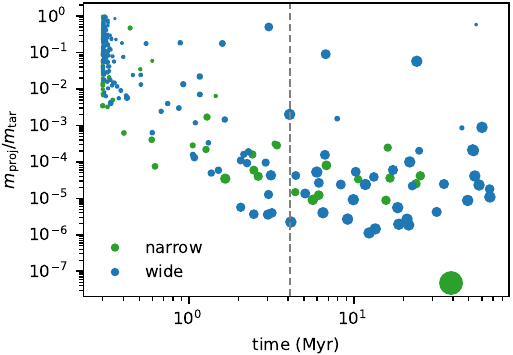}}
\caption{Collisions as a function of time. We show the projectile-to-target mass ratio for each collision for runs \texttt{n05} (green) and \texttt{w05} (blue) from $0.3$ to $100\,\mathrm{Myr}$. Each symbol shows a single collision. The nebular phase ends at $4.1\,\mathrm{Myr}$ (dashed line). The symbol size scales with the mass of the target as ${\propto}m^{1/3}$.}
\label{fig:collisions}
\end{figure}
Most collisions occur in the first $0.1\,\mathrm{Myr}$ when the surface densities of planetesimals are high. As time goes on, collisions become less frequent. Figure~\ref{fig:collisions} shows the projectile-to-target ratio as a function of time for the narrow and wide ring runs \texttt{n05} and \texttt{w05} from $0.3$ to $100\,\mathrm{Myr}$. Initially, the mass ratio of projectile and target is ${\gtrsim}10^{-2}$ and most collisions are between planetesimals or between the top of the mass distribution and planetesimals. At later times and after the disc dissipated, the mass ratio of projectile and target decreases with time and is typically ${\lesssim}10^{-2}$. This indicates that collisions are dominantly between a planet-sized body and a much smaller planetesimal. Giant impacts between planetary cores with a mass ratio ${\gtrsim}0.1$ appear to be typically rare. However, we integrated for only $100\,\mathrm{Myr}$, on longer $\mathrm{Gyr}$ timescales this may change and we leave this for future work.

Generally, giant impacts appear to be rare in the giant-planet formation region within the first $100\,\mathrm{Myr}$ of evolution. In the Solar System, giant impacts are linked to the Moon-forming impact ($t{\approx}50{-}100\,\mathrm{Myr}$; \citealt{Canup2004,Thiemens2019}), and potentially to the obliquities of Uranus and Neptune (\citealt{Morbidelli2012,Izidoro2015,Kegerreis2018,Reinhardt2020,Esteves2025}, but see also e.g. \citealt{Boue2010} and \citealt{Saillenfest2022} for a non-impact mechanism). It is challenging to place our results in the context of giant impacts in the Solar System, as we do not model terrestrial planet formation, and giant impact probabilities, conditional on the presence of four giant planets, would require a much larger simulation suite than presented in this work.

\subsection{Mass distribution}

\begin{figure}
\resizebox{\hsize}{!}{\includegraphics{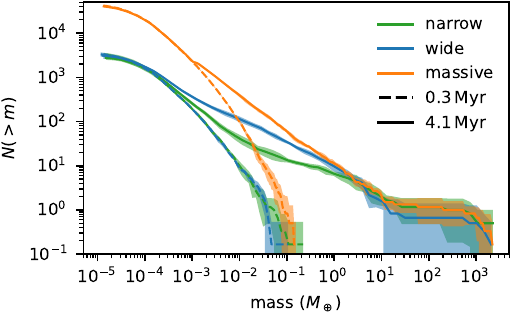}}
\caption{Mass distribution at $4.1\,\mathrm{Myr}$. We show the average mass distributions for the narrow (green), wide (blue), and high-mass wide (orange) rings at $4.1\,\mathrm{Myr}$ (solid lines) with respective standard deviation (shaded area). The initial distributions (dashed lines) are shown for comparison.}
\label{fig:finalccdf}
\end{figure}
Figure~\ref{fig:finalccdf} shows the mean complementary cumulative mass distribution, that is the number of bodies larger than mass $m$, at the end of the nebular phase at $4.1\,\mathrm{Myr}$ and, for comparison, the initial distribution for each set of simulations (Table~\ref{tab:initialrings}). For all three cases, the mass distributions are similar for planetesimals (${<}10^{-3}\,M_\oplus$) and resemble the initial distribution showing that neither collisions nor pebble accretion result in significant growth of this population. For planets more massive than ${\sim}1{-}2\,M_\oplus$, the distributions are comparable as well. This indicates that neither the initial distribution nor the total mass in planetesimals significantly impact the final number of planets.

However, in the low-mass runs, bodies in the mass range ${\sim}10^{-3}{-}1\,M_\oplus$ are depleted by up to a factor ${\sim}4$ in the narrow ring runs compared to the wide rings. This is a consequence of the reduction of pebble accretion efficiency due to dynamical excitation of the pebble accreting bodies \citep{Levison2015,Lau2024}. Because of the higher spatial concentration of planetesimals in the narrow rings compared to the wide rings, encounter rates are higher and planetesimal scattering results in higher eccentricities and inclinations. Therefore, the relative velocities between pebbles and planetesimals are higher for the narrow rings, especially in earlier times before high-mass planets dominate the scattering and erase the effect of the initial conditions. Because our pebble accretion prescription accounts for the relative velocity between the accreting body and the pebbles, higher relative velocities lead to shorter encounter times, and consequently, the weak-coupling accretion rate applies, which lowers the efficiency for pebble accretion. For higher mass planets, eccentricity and inclination damping circularises their orbits and this effect vanishes. This explains why bodies in the mass range ${\sim}10^{-3}{-}1\,M_\oplus$ grow less efficiently in the narrow rings. 

We ran a set of simulations with an initially higher total mass of planetesimals of ${\sim}10\,M_\oplus$ (runs \texttt{m01} to \texttt{m06}) to study whether this affects the final mass distribution of planets. Comparing the high-mass wide rings with the low-mass wide rings (Fig.~\ref{fig:finalccdf}), shows that the number of planets ${\gtrsim}1\,M_\oplus$ are comparable, whereas the number of lower-mass bodies is higher. The higher number of lower-mass bodies is a natural consequence of the higher number of planetesimal seeds. The self-regulation of the number of ${\gtrsim}1\,M_\oplus$ planets is due to mutual filtering, which we discuss in more detail in Sect.~\ref{sec:reductionofpebbleflux}.

\subsection{Final configuration of planets}

\begin{figure}
\centering
\resizebox{\hsize}{!}{\includegraphics{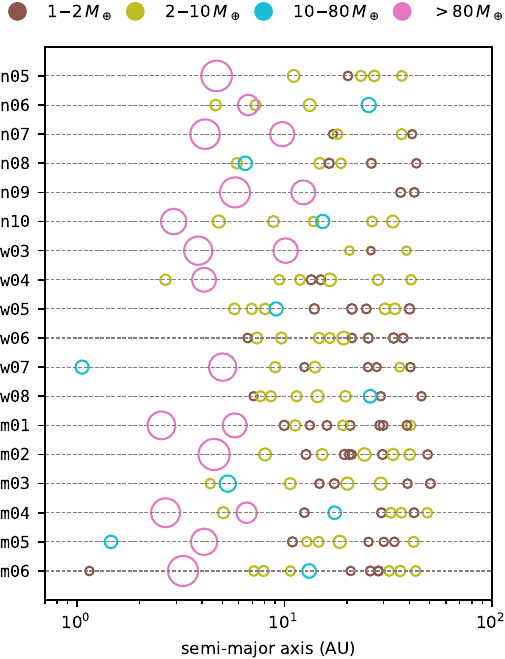}}
\caption{Configuration of planets at $4.1\,\mathrm{Myr}$. Each row shows the configuration of planets ${\ge}1\,M_\oplus$ at $4.1\,\mathrm{Myr}$. The runs we show are as follows: narrow ring runs (\texttt{n05}{-}\texttt{n10}, top), wide ring runs (\texttt{w03}{-}\texttt{w08}, middle), and high-mass ring runs (\texttt{m01}{-}\texttt{m06}, bottom). The symbol size scales with the mass of the body as ${\propto}m^{1/3}$, and the colour indicates the mass category.
}
\label{fig:architecture}
\end{figure}
The final configuration of planets for all simulation runs is summarised in Fig.~\ref{fig:architecture}. We start with the configuration of the planets. In all our simulations we find that the most massive planets, that is those that evolve into gas giants, migrate inwards and end up in the range ${\sim}2{-}10\,\mathrm{AU}$. Lower-mass planets, that is those in the terrestrial to super-Earth range, are predominantly found outside the orbit of the gas giants with semi-major axis ${\sim}6{-}50\,\mathrm{AU}$. These planets form and remain closely to the initial location of the planetesimal disc. The same holds for ice giants, even though in two runs (\texttt{w07} and \texttt{m05}), one ice giant respectively ends up in the range $1{-}2\,\mathrm{AU}$, inside the orbit of the giant planet. During formation, these two ice giants got dragged along inwards by the giant planet and continued to migrate. The same occurred in run \texttt{m06} for the terrestrial-like planet that ended up close to $1\,\mathrm{AU}$.

\begin{figure}
\centering
\resizebox{\hsize}{!}{\includegraphics{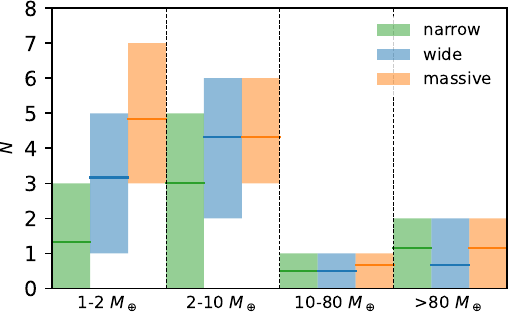}}
\caption{Number of planets ${\ge}1\,M_\oplus$ at $4.1\,\mathrm{Myr}$. We show the min-max number range (shaded) and the mean number (solid line) of planets formed over all simulations for the narrow (green), wide (blue), and high-mass wide (orange) rings.
}
\label{fig:planetstat}
\end{figure}
Figure~\ref{fig:planetstat} summarises how many planets in each mass category ${\ge}1\,M_\oplus$ form in our simulations. Ice and gas giants are typically rare. Regardless of the initial distribution and initial total mass of planetesimals, at most two gas giants are formed and at most one ice giant per run. The rare occurrence of ice giants may be a consequence of the pebble isolation mass falling in the same mass range in the $10{-}50\,\mathrm{AU}$ range. Therefore, if planets grow too slowly, they will end up with masses less than $10\,M_\oplus$, and, on the other hand, if they grow more quickly and reach pebble isolation mass, they will continue to grow to gas giants by rapid gas accretion. In the mass range $2{-}10\,M_\oplus$, roughly the same number of ${\sim}2{-}6$ planets are formed in the low- and high-mass wide ring runs. The narrow rings show a higher variation because run \texttt{n09} did not form any planets in this mass range. Excluding this specific run, however, shows that also the remaining narrow ring runs produce $2{-}5$ planets in the super-Earth regime. Finally, going to planets ${\lesssim}2\,M_\oplus$, we find that the narrow rings form the lowest number of planets in this mass range. Typically, we find one to three planets, excluding runs \texttt{n06} and \texttt{n10} that did not form any planets in this mass range.

The wide ring runs form one to five planets, and the high-mass wide ring runs form three to seven planets in this range. It is worth noting that, based on these numbers, the initial configuration and the initial total mass of the planetesimals have only minor influence on the planetary systems formed.

\subsection{Reduction of pebble flux}
\label{sec:reductionofpebbleflux}

\begin{figure}
\centering
\resizebox{\hsize}{!}{\includegraphics{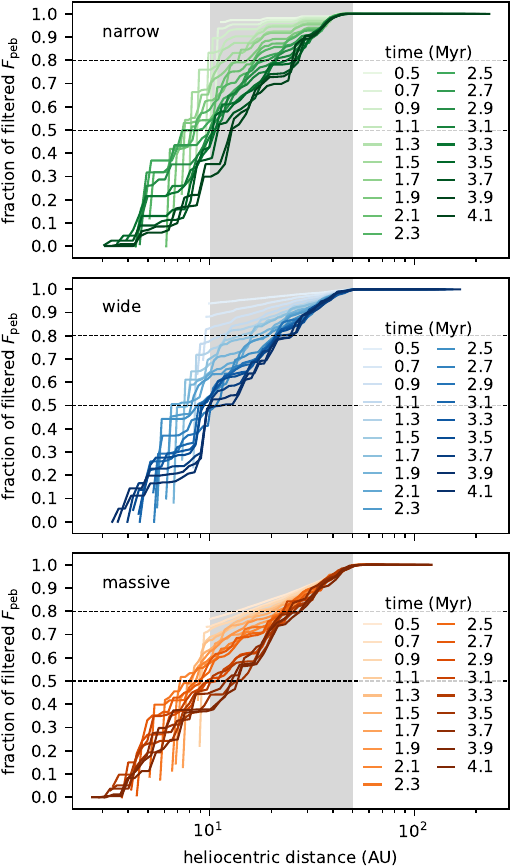}}
\caption{Reduction of the pebble flux through accreting planets. Each line shows the pebble flux in a given heliocentric distance bin at a given time averaged over all simulations for an initial configuration of planetesimals. The panels are as follows: narrow rings (top), wide rings (middle), and high-mass wide rings (bottom). The initial location of the planetesimals (grey area) and the $50\,\%$ and $80\,\%$ reduction levels (dashed lines) are shown for reference.
}
\label{fig:filtering}
\end{figure}
In Fig.~\ref{fig:filtering} we show the reduction of the pebble flux through concurrently accreting bodies in our simulations to investigate the effect of initial distribution and initial total mass of planetesimals. To do so, we show the pebble flux in a given heliocentric distance bin at different times. Because $N$-body simulations are stochastic, and the reduction depends on the configuration and number of efficiently accreting bodies at a given time, we averaged over all runs for a given set of simulations to identify the general behaviour.

Starting at large heliocentric distances, we see that the pebble flux reduces gradually towards the inner disc. This is expected because the number of bodies that accrete pebbles increases reducing the flux of pebbles through the system. The reduction is smoother at early times compared to later times, where we see steps in the pebble flux. This is a consequence of the planetesimals growing to embryos and planets. Initially, all planetesimals are small and confined to their initial location, accreting a similar fraction of pebbles. At later times, more embryos emerge that dominate the accretion of pebbles and hence the reduction of the pebble flux. While the embryos grow and migrate inward, the pebble flux to the inner disc reduces. Because we assume that a planet reaching pebble isolation mass cuts the flux of pebbles and because these massive planets migrate swiftly, the pebble flux goes to zero at small heliocentric distances.

The initial configuration of planetesimals has a minor effect on the final pebble flux reduction. The reduction with decreasing heliocentric distance is more gradual for the wide ring runs. This is a consequence of the spread-out configuration of the planetesimals where bodies at all heliocentric distances contribute to the flux reduction. In comparison, pebbles are accreted stepwise at the ring locations for the narrow rings before the distribution widens due to dynamical excitation. Furthermore, the dynamically more excited bodies in the narrow rings have lower accretion efficiencies, therefore slightly lowering the total reduction in a given distance bin at a given time. The pebble flux is reduced to ${\sim}80\,\%$ at ${\sim}10\,\mathrm{AU}$ in ${\lesssim}1\,\mathrm{Myr}$ for the wide rings, whereas the same level of reduction is reached after ${\sim}1\,\mathrm{Myr}$ to $1.5\,\mathrm{Myr}$ for the narrow rings. By $4.1\,\mathrm{Myr}$, the reduction of the pebble flux close to ${\sim}10\,\mathrm{AU}$ reaches ${\sim}50\,\%$.

The total initial mass of planetesimals results in a stronger reduction of the flux at early times. Towards the end of the simulation, the reduction reaches ${\sim}50\,\%$ at ${\sim}10\,\mathrm{AU}$ as in the runs with lower initial total mass in planetesimals. However, the higher mass in planetesimals, and therefore the higher number of accreting bodies results in a significant reduction of the pebble flux to ${\lesssim}80\,\%$ at ${\sim}10\,\mathrm{AU}$ already at ${\lesssim}1\,\mathrm{Myr}$.

\section{Discussion}
\label{sec:discussion}

\subsection{Model limitations}

\subsubsection{Pebble accretion model}
Pebble accretion depends significantly on the chosen model for the protoplanetary disc because the scale height of the disc is an important parameter affecting the mass accretion rates of the embryos. The scale height is directly linked to the temperature profile of the disc, which we assumed here to be set by stellar irradiation and is a valid assumption for the outer parts of the disc. However, including accretional heating may have a significant influence on suppressing pebble accretion in the inner part of disc, which may become important for the inwards migrating embryos \citep{Danti2025}.

Pebble accretion depends on the size of the pebbles \citep{Lyra2023}. However, we do not include coagulation or fragmentation in our model to obtain the full size distribution of pebbles and instead use a single representative size given by the limiting cases in Sect.~\ref{sec:pebblestokesnumber}. We assume that upon reaching pebble isolation mass, a planet cuts off the pebble flux, whereas studies of dust growth and transport through planet-induced gaps shows that small dust grains may leak through the gap \citep{Stammler2023}. However, planets that reach pebble isolation mass in our simulations typically originate close to the inner edge of the initial planetesimal distribution or migrate inwards by type-I migration and therefore do not affect growth by pebble accretion in the outer disc.

\subsubsection{Gas accretion}
The gas accretion prescription chosen in our work is simplified. While the first stage of Kelvin-Helmholtz contraction is a well-established concept \citep{Ikoma2000}, the second stage of runaway gas accretion is merely a power-law fit to numerical simulations \citep{Lambrechts2019b} that mimics the actual amount of gas entering the Hill sphere of the planet that is accreted. Compared to more detailed prescriptions where the gas accretion rate is regulated by the gap opened by the planet \citep{Tanigawa2016,Ida2018}, our prescription results in higher gas accretion rates. Consequently, the transition from fast type-I to slower type-II migration occurs earlier in our simulations, which may explain why the gas giants in our simulations remain outside $1\,\mathrm{AU}$ compared to previous work \citep[e.g.][and discussion in Sec.~\ref{sec:comparison}]{Bitsch2019,Lau2024}.

\subsubsection{Numerical limitation}
The GPU-accelerated $N$-body methods allow significantly more fully interacting bodies to be included in planet formation simulations compared to conventional CPU-based methods. In our nominal simulations, we included around $4000$ fully interacting bodies ranging from planetesimals of a few hundred kilometres to approximately $0.1\,M_\oplus$ embryos for the most massive bodies. With this setup, the nebular phase of $4.1\,\mathrm{Myr}$ takes a few days to complete and the gas-free phase up to a week. Including more bodies, such as in the high-mass runs with approximately $40\,000$ bodies, the computational time increases drastically to weeks, even though only ${\sim}4000$ bodies are fully interacting and the rest are treated as test particles. The most time-consuming part in our simulations is the reduction of the pebble flux, which requires sorting of all pebble accreting bodies in the simulation. We leave an optimisation of this issue to future work.

\subsection{Comparison with previous work}
\label{sec:comparison}

Previous $N$-body work on planet formation by pebble accretion focused on growth of a limited number of embryos to study the effect of disc parameters, pebble flux, or pebble accretion models on the final configuration of planets. Other studies focused specifically on understanding the formation of the giant planets in the Solar System and consequences on for example, terrestrial planet formation or mixing of asteroidal material.

\citet{Levison2015} pioneered the use of $N$-body simulations of pebble accretion to study the growth of gas giants in the Solar System. Their model showed the formation of one to four gas giants in the $5{-}15\,\mathrm{AU}$ range. This number and heliocentric distance range is comparable to our study, even though \citet{Levison2015} neglected migration, whereas in our model the gas giants migrate from initially ${\gtrsim}10\,\mathrm{AU}$ to their final locations between $3$ and $15\,\mathrm{AU}$. This shows that despite differences in initial conditions and model parameters, pebble accretion is a robust mechanism to form giant planets. Furthermore, \citet{Levison2015} pointed out the effect of dynamical excitation of embryos on their growth rates. Excitation slows down growth by increasing the relative velocities between the accreting body and the pebbles, which reduces the pebble accretion efficiency, and by increasing the inclination of the embryos, which leads to reduced accretion efficiency because the embryos spend a significant amount of time above or below the pebble scale height. We see both effects in our simulations as well.

\citet{Bitsch2019} use $N$-body simulations of pebble accretion to explore under which conditions embryos grow to gas giants. They present simulations with $60$ embryos in the mass range $0.005$ to $0.015\,M_\oplus$ distributed between $10$ and $40\,\mathrm{AU}$, which is similar to our initial configuration, even though we used a full mass distribution of planetesimals instead of embryos. Their scaling factor $S_\mathrm{peb}{=}2.5$ for the pebble surface density corresponds to a total pebble mass of ${\sim}175\,M_\oplus$ drifting through the gas disc during its lifetime of $3\,\mathrm{Myr}$. In our simulations, the total mass of pebbles integrated from $0.3$ to $4.1\,\mathrm{Myr}$ is ${\sim}163\,M_\oplus$, which is comparable. For this pebble flux, \citet{Bitsch2019} observe that only embryos located around $10\,\mathrm{AU}$ grow fast enough to eventually accrete gas, but migration is too fast and places those planets close to ${\sim}1\,\mathrm{AU}$. Embryos farther out than ${\sim}15\,\mathrm{AU}$ do not grow quickly enough to reach the threshold for runaway gas accretion to start. For higher pebble fluxes or slower migration speeds, \citet{Bitsch2019} show that embryos grow fast enough to initiate gas accretion and migrate sufficiently slow to remain outside $1\,\mathrm{AU}$.

\citet{Raorane2024} explore the formation of ice and gas giant planets in the Solar System with $N$-body simulations of pebble accretion. On average, \citet{Raorane2024} form $1.7$ planets ${\gtrsim}10\,M_\oplus$ with an average multiplicity of $2.5$. This is broadly consistent with our simulations. Furthermore, \citet{Raorane2024} find that ice-giant analogues are rare and that the majority of giant planets exceeds Saturn in mass. Also this is consistent with our simulations where we see a similar trend.

\citet{Matsumura2017} and \citet{Matsumura2021} perform global $N$-body simulations of pebble accretion to study the influence of various disc parameters, such as dissipation timescale, disc mass and metallicity, on the formed planets. They run $240$ simulations for different parameters in which they follow the growth and dynamics of $10$ cores with initial mass of $10^{-2}\,M_\oplus$ distributed over $0.5{-}15\,\mathrm{AU}$. \citet{Matsumura2021} successfully reproduce the overall distribution trends of semi-major axis, eccentricity, and mass of extrasolar giant planets. Similar to results presented here, they point out that the formation of gas giants is common when the core growth time is shorter than the gas disc dissipation time, else lower-mass ice giants form \citep{Lambrechts2014b}.

\citet{Lau2024} study planet formation starting from planetesimals using $N$-body simulations in a similar fashion as in this work. They confirm that dynamical excitation suppresses pebble accretion because encounter times between pebbles and the accreting body become short \citep{Levison2015}. We see the same effect in our simulations where growth by pebble accretion of bodies in the ${\sim}10^{-3}{-}1\,M_\oplus$ is suppressed in the dynamically more excited narrow ring runs compared to the dynamically colder wide ring runs. Furthermore, \citet{Lau2024} find that typically one to two gas giants and one to two ice giants form in their simulations outside ${\sim}6\,\mathrm{AU}$ if migration is not included. However, with migration, massive cores migrate out of their simulation domain before gas accretion could set in resulting in no giant planets. \citet{Lau2024} use a different prescription for gas accretion, planet-disc interactions, and the transition mass to gap-induced type-II migration \citep{Ida2018}. Our gas accretion prescription \citep{Lambrechts2019b} has higher gas accretion rates, and consequently gas giants transition earlier from fast type-I to slow type-II migration, thus preventing the giant planets from being lost to the inner disc.

\subsection{Implications for cold giant exoplanets}

It is outside the scope of this work to perform population synthesis and do a statistical comparison to the exoplanet census, which would require a substantially larger simulation suit to explore the effects of disc diversity and to trace stochastic variability \citep{Matsumura2021}. 
Instead, we briefly comment on some general findings.

Firstly, in systems where gas giants form, we find that they typically orbit between ${\sim}3$ to ${\sim}10\,\mathrm{AU}$, and not outside of $20\,\mathrm{AU}$ (see Figs.~\ref{fig:architecture}). This appears to be broadly in line with an inferred turn over in giant plant occurrence rates at wider orbits outside of ${\sim}5\,\mathrm{AU}$ \citep{Fulton2021} and the low occurrence of super-Jupiters outside of $20\,\mathrm{AU}$, as inferred from direct imaging surveys \citep{Bowler2018,Vigan2021}. We also find that, in general, giant planets form in multiples, which appears consistent with inferred giant exoplanet multiplicities \citep{Rosenthal2024}. However, the number of specifically gas giants does not frequently exceed two regardless of initial distribution and initial total mass of planetesimals.

Finally, we briefly comment on the low eccentricities of the giant planets in our simulations, that are typically low and comparable to the eccentricities of the giant planets in the Solar System with $e{\lesssim}0.1$ (see Fig.~\ref{fig:snapnarrowgf}). In contrast, the eccentricities of observed exoplanets are typically high ($\gtrsim0.1$), which may be attributed to planet-planet scattering or planet-disc interactions \citep[e.g.][]{Dawson2013,Buchhave2018,Bitsch2020b}. The here obtained eccentricities may depend on the employed eccentricity damping prescription, as explored in \citet{Bitsch2020b}, and would benefit from explicit hydrodynamical modelling \citep{Griveaud2024}.

\subsection{Implications for inner planets}

We investigated the formation of planets in the outer disc, starting with planets spread out between $10{-}50\,\mathrm{AU}$. The reduction of the pebble flux through concurrently accreting embryos, can reduce the pebble flux interior to $10\,\mathrm{AU}$ by up to $40\,\%$ within the first $\mathrm{Myr}$ of evolution, depending on the total initial mass of planetesimals. This may have implications for the inner disc, where terrestrial planets form \citep{Lambrechts2019}. \citet{Danti2025} show that the reduction of the pebble flux by an outer planet alone is typically not sufficient to suppress the growth of inner embryos, unless the inner disc is also strongly heated by accretion heating in the midplane, which increases the gas scale height and suppresses inside-out pebble accretion \citep{Chachan2023,Yap2024}.

\subsection{Scattered disc of not-so-minor bodies}

\begin{figure}
\centering
\resizebox{\hsize}{!}{\includegraphics{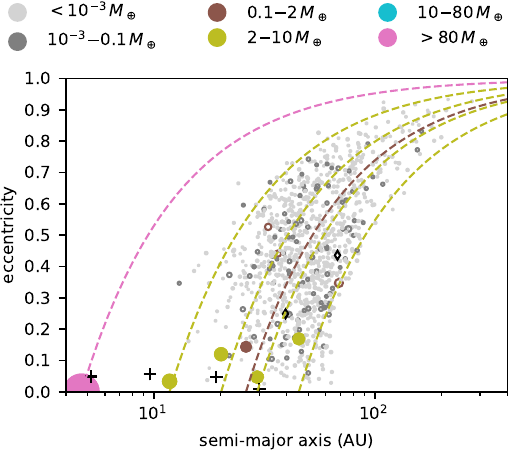}}
\caption{
Distribution of planetesimals at $100\,\mathrm{Myr}$ for run \texttt{n05}. We show the eccentricity curves where perihelion equals the semi-major axis of the corresponding planets (dashed lines) for reference. The symbol size scales with the mass of the body as ${\propto}m^{1/3}$ and planets ${\ge}1\,M_\oplus$ are marked as solid circles.
}
\label{fig:scattered}
\end{figure}
Our simulations show the generic production of a scattered disc of planetesimals and embryos originating from the giant-planet-forming zone. These bodies, with semi-major axis up to $100{-}200\,\mathrm{AU}$, have a wide mass distribution: from planetesimals originating from the low-mass end of the primordial planetesimal distribution, to Pluto-sized embryos, and onwards to a lower number of planets exceeding the Earth in mass that underwent pebble accretion (Fig.~\ref{fig:scattered}). Such discs may be connected to frequently observed debris discs around young stars. The inferred debris and dust production rates are difficult to generate though self-stirring without exceeding the solid mass reservoir of typical protoplanetary disc \citep{Krivov2021}. It has therefore been suggested that the required stirring rates could be connected to the presence of nearby planets \citep{Mustill2009,Munoz2023,Costa2024}. Recent observations support at least some of these discs to host a scattered population of bodies: HR~8799 \citep{Geiler2019}, $\beta$~Pictoris \citep{Matra2019}, and TWA~7 \citep{Lagrange2025}. The presence of massive embryos in the scattered disc may be a key ingredient to explain debris disc collision rates \citep{Costa2024}. 

Although our simulations should not be seen in the context of studies exploring the origin and dynamical sculpting of the Kuiper belt, the natural outcome of a wide mass distribution of scattered objects does support recent explorations that invoke the presence of a substantial population of Pluto-mass objects \citep{Nesvorny2016b}, potentially more massive objects \citep{Huang2022}, and opens the avenue to implant objects in the Earth-to-ice giant mass regime \citep{Batygin2019}. Since our simulations only trace the initial planetesimal population, making a direct comparison to the Kuiper belt is challenging. Moreover, to assess whether our generic initial conditions could naturally lead to a true outer Solar-System analogue, we would require a larger suite of simulations to trace Solar-System-like dynamical instabilities among giant planets \citep{Gomes2005,Tsiganis2005,Morbidelli2010,Griveaud2024}. However, while the dynamically hot population and the scattered disc were implanted by scattering \citep{Duncan1997,Levison2008,Nesvorny2016b}, the cold classical Kuiper belt binaries are evidence for in situ planetesimal formation by streaming instability \citep{Nesvorny2019}, possibly formed late in the evolution of the gas disc \citep{Carrera2017,Ercolano2017,Lau2025}.

\section{Conclusion}
\label{sec:conclusions}

In this work, we have conducted GPU-accelerated $N$-body simulations of planet formation with pebble accretion using \texttt{GENGA} \citep{Grimm2014,Grimm2022}. We placed the location of our study in the giant-planet formation region between $10$ and $50\,\mathrm{AU}$. Starting with a streaming instability inspired initial mass distribution of planetesimals, we simulated $4.1\,\mathrm{Myr}$ of evolution in the nebular phase and extended the simulations to $100\,\mathrm{Myr}$ in the gas-free phase. We tested two scenarios for the initial radial distribution of planetesimals. In the narrow ring runs, planetesimals were distributed in localised narrow rings between $10$ and $50\,\mathrm{AU}$, where each ring had the typical width of a streaming instability filament. In the wide ring runs, the planetesimals were distributed such that the rings connect to represent a more spatially uniform distribution of bodies. In each case, we used a total mass of $1\,M_\oplus$ in planetesimals. For comparison, we also conducted a set of wide ring simulations where we used a ten times higher total mass of planetesimals of $10\,M_\oplus$. Our main findings are summarised as follows.
\begin{itemize}
    \item The initial spatial distribution of planetesimals does not significantly affect the final outcome of our simulations. Starting with narrow rings, scattering and diffusion of planetesimals efficiently redistributes these bodies within the first megayear. In this sense, planetary architectures do not retain a memory of the initial planetesimal distribution.
    \item The qualitative evolution and the final outcome of the narrow and wide ring runs are similar. We consistently find that one to two gas giants form, paired with a handful of icy planets exceeding two Earth masses (Fig 10).
    \item The initial total mass of planetesimals does not significantly affect the number of planets that form with masses ${\gtrsim}1{-}2\,M_\oplus$. Reduction of the pebble flux by a high number of concurrently accreting bodies results in a self-regulated growth. This counteracts the expected higher number of planets due to the increased number of seeds resulting in the formation of a similar number of gas giants in the high-mass wide ring simulations with initially $10\,M_\oplus$ of planetesimals compared to the nominal runs with initially only $1\,M_\oplus$.
    \item Our simulations argue that the formation of a scattered disc of planetesimals outside of ${\sim}20\,\mathrm{AU}$, with planets with masses ${\gtrsim}0.1\,M_\oplus$ embedded in it, is a natural outcome of giant-planet core formation.
    \item Giant impacts between planetary cores generally appear to be rare in the first $100\,\mathrm{Myr}$.
\end{itemize}

\begin{acknowledgements}
S.~Lorek and M.~Lambrechts acknowledge this work is funded by the European Research Council (ERC Starting Grant 101041466-EXODOSS). The Tycho supercomputer hosted at the SCIENCE HPC center at the University of Copenhagen was used for supporting this work. We thank Simon Grimm for helpful discussions during the preparation of this work. We thank the anonymous referee for the constructive report that helped to improve the quality of this manuscript.
\end{acknowledgements}

\bibliographystyle{aa}
\bibliography{ref}

\begin{appendix}

\end{appendix}

\end{document}